\documentclass[preprint,aps,pra,showpacs,floatfix]{revtex4}
\usepackage{amsmath}
\usepackage{times}
\usepackage{euscript}
\usepackage{graphicx}
\usepackage{amsthm}
\usepackage{epsf}
\usepackage{epsfig}
\usepackage{bm}
\usepackage{soul}
\usepackage{xcolor}
\graphicspath{ {./figs/} }
\renewcommand{\vec}[1]{{\mbox{\boldmath$#1$}}}

\newcommand*\rfrac[2]{{}^{#1}\!/_{#2}}
\usepackage[linktocpage=true,plainpages=false,pdfpagelabels=false]{hyperref}
\begin{document}
\thispagestyle{empty}
\title{Electron-positron pair production in slow
collisions of heavy nuclei beyond the monopole approximation}
\author{I.~A.~Maltsev,$^{1}$
V.~M.~Shabaev,$^{1}$
R.~V.~Popov,$^{1}$
Y.~S.~Kozhedub,$^{1, 2}$
G.~Plunien,$^{3}$
X.~Ma,$^{4}$
and Th.~St\"ohlker$^{5,6,7}$
}
\affiliation{
$^1$ Department of Physics, St. Petersburg State University,
Universitetskaya naberezhnaya 7/9, 199034 St. Petersburg, Russia\\
$^2$ NRC “Kurchatov Institute”, Academician Kurchatov 1, 123182 Moscow, Russia\\
$^3$ Institut f\"ur Theoretische Physik, Technische Universit\"at Dresden,
Mommsenstra{\ss}e 13, D-01062 Dresden, Germany\\
$^4$
Institute of Modern Physics, Chinese Academy of Sciences, 
730000  Lanzhou, China\\
$^5$
GSI Helmholtzzentrum f\"ur Schwerionenforschung GmbH,
Planckstrasse 1, D-64291 Darmstadt, Germany \\
$^6$Helmholtz-Institute Jena, D-07743 Jena, Germany\\
$^7$Theoretisch-Physikalisches Institut,
Friedrich-Schiller-Universit\"at Jena, D-07743 Jena, Germany
\vspace{10mm}
}
%
\begin{abstract}
Electron-positron pair production in low-energy collisions of heavy nuclei 
is considered beyond the monopole approximation.  The calculation method is 
based on the numerical solving of the time-dependent 
Dirac equation with the full two-center potential. 
Bound-free and free-free pair-production probabilities as well 
as the energy spectra of the emitted positrons are calculated
for the collisions of bare uranium nuclei.  
The calculations are performed for collision energy near the Coulomb barrier
for different values of the impact parameter. 
The obtained results are compared with the corresponding values calculated
in the monopole approximation.
\end{abstract}
\maketitle
\section{INTRODUCTION}
Heavy quasimolecules formed in low-energy ion collisions provide a unique opportunity 
to study quantum electrodynamics~(QED) in extremely strong fields. The 
ground state of the quasimolecule with the total nuclear charge exceeding 
the critical value~$Z_{\rm cr} \approx 173$ can dive into the 
negative-energy Dirac 
continuum~\cite{Pomeranchuk_45, Gershtein_70, 
Greiner_69, Zeldovich_71, Greiner_85, Mueller_94, Godunov_17}. 
As was predicted in Refs.~\cite{Gershtein_70, Greiner_69}, after diving the ground 
state appears as a resonance which can decay spontaneously with emission of a positron. 
The detection of the emitted particles would confirm 
QED theory in the unexplored supercritical regime. 
However, the dynamics of the nuclei can also induce pair production. 
Therefore the detection of the supercritical resonance decay requires distinction of 
spontaneous and dynamical contributions. 

The early experimental investigations of supercritical
heavy-ion collisions were performed in GSI (Darmstadt, Germany). But no evidence of 
the diving phenomenon was found~\cite{Mueller_94}. The next generation of accelerator
facilities is expected to drive these investigations to 
a new level~\cite{Gumbaridze_09, Ter-Akopian_15, Ma_17}.  
The experimental study requires the proper theoretical analysis. 

The first calculations of 
pair production in supercritical collisions were based on the quasistationary approach
in which only the spontaneous contribution was taken into account~\cite{Gershtein_73, Popov_73, Peitz_73}.
The dynamical pair production in low-energy ion collisions was investigated in Ref.~\cite{Lee_16} 
using the perturbation theory. However, this approach is restricted to the relatively small values of 
the nuclear charge and cannot be applied to the supercritical case. 
A rough analytical estimation of pair-production cross section for 
heavy nuclei was done in Refs.~\cite{Khriplovich_14, Khriplovich_16, Khriplovich_17}.
The dynamical methods employ the solving 
of the time-dependent Dirac equation~(TDDE) 
which can be performed numerically using
various techniques~\cite{Reinhardt_81, Muller_88, Ackad_07, Ackad_08, Maltsev_15, 
Bondarev_15, Maltsev_17, RV_Popov_18}. 
These methods take into account the dynamical pair-production mechanism as well as
the spontaneous one.
{
However, there is no direct way to distinguish their contributions to the obtained results.
The influence of each mechanism can be investigated only via comparison 
of the values calculated for the subcritical and supercritical collisions. 
Since in the subcritical case there is no diving, the pair production should be of pure dynamical origin. 
In the calculations, the existence of the spontaneous mechanism was demonstrated either 
via introducing the sticking of the nuclei at the closest approach~\cite{Reinhardt_81, Muller_88, Maltsev_15} 
or via slowing them artificially down~\cite{Maltsev_15}. In both 
scenarios, one could see the enhancement of the pair production in the supercritical case  
that can be explained only with the spontaneous mechanism.  But  the experimental
realization of these scenarios is very questionable. 
}

{One can try to find the signal from the
supercritical resonance decay in the differential characteristics of the created particles
such as positron energy distribution. 
But, as was shown by Frankfurt
group~(see~Refs.~\cite{Greiner_85, Reinhardt_81, Muller_88} and references therein) 
and recently confirmed in Ref.~\cite{Maltsev_15}, 
this signal cannot be found in the positron spectra of the elastic collisions due to the 
dominant role of the dynamical pair creation. 
However, these calculations were performed within the monopole approximation 
where only the spherically symmetric part of the two-center potential is considered.

It is known that the binding energies of the lowest quasimolecular states
calculated in the monopole approximation are in rather good agreement with the exact 
two-center ones at short internuclear distances~\cite{Greiner_85, Mueller_94, Tup_10}. 
Since this region seems to be the most important for the pair production, 
one can assume that it can be quite well described with the monopole part of the
two-center potential. However, the influence of effects beyond the 
monopole approximation on electron dynamics, which determines the pair production, 
cannot be estimated without two-center time-dependent calculations.
Therefore, in order to verify the obtained monopole results, it is 
necessary to perform the related calculations which require the corresponding 
theoretical methods.

Such methods were developed in 
Refs.~\cite{Maltsev_17, RV_Popov_18}, where they were applied to calculations of the total 
bound-free pair-production probabilities. 
But, in order to investigate the possibility of 
the observation of the diving phenomenon in the same manner as in the monopole case, 
one needs to calculate the energy spectra
of the emitted positrons. It is also necessary to take into account
the free-free pair production which also contributes to the spectra. 
In the present work, we have performed the required calculations 
using a method similar to Ref.~\cite{Maltsev_17}. }

The approach is based on the time 
evolution of the finite number of initial one-electron 
states via numerical solving of the TDDE with the 
full two-center potential. The TDDE is considered in the coordinate frame rotating 
with the internuclear axis. The time-dependent electron 
wave functions are expanded in a finite basis set 
constructed from B-splines using the dual-kinetic balance 
approach for axially symmetric systems~\cite{Rozenbaum_14}. 
The pair-creation probabilities are obtained 
utilizing the expressions known from QED theory 
with unstable vacuum~\cite{Greiner_85, Fradkin_91}.

Throughout the paper we assume $\hbar = 1$.

\section{THEORY}
\label{sec:theory}
\subsection{Pair production}
\label{sub:pair_production}
In the present work, we consider the interaction of electrons 
with the strong external electromagnetic field 
nonpertubatively but neglect the interelectronic
interaction as well as the interaction with the quantized radiation field.
The electron dynamics in presence of the external field is governed by the time-dependent 
Dirac equation:
\begin{equation}
 i \frac{\partial}{\partial t} \psi (\vec{r}, t)  = 
 {H}_D (t) \, \psi (\vec{r}, t) \, ,
 \label{eq:time_dirac}
\end{equation}
\begin{equation}
{H}_D (t)= c \vec{\alpha} \left( \vec{p} - e\vec{A}(\vec r ,t)\right)+V(\vec r, t)+ \beta m_e c^2.
\end{equation}
Here $(\vec{A}, V)$ describe the interaction with the external field, 
$(\vec \alpha, \beta)$ are the Dirac matrices, $m_e$ is the electron mass, 
$e < 0$ is the electron charge, and $c$ is the speed of light. 
Let us introduce two sets of solutions of Eq.~\eqref{eq:time_dirac} with
the different asymptotics:
\begin{equation}
 \psi^{(+)}_i(\vec{r}, t_{\rm in})=\varphi^{\rm in}_i (\vec{r}), \,
 \psi^{(-)}_i(\vec{r}, t_{\rm out})=\varphi^{\rm out}_i (\vec{r}),
\end{equation}
where $t_{\rm in}$ is the initial and $t_{\rm out}$ is the final time moment,
and $\varphi^{\rm in}_i (\vec{r})$ and $\varphi^{\rm out}_i (\vec{r})$ 
are the eigenfunctions of corresponding instantaneous Hamiltonians: 
\begin{equation}
 {H}_D(t_{\rm in}) \, \varphi^{\rm in}_i (\vec{r})=\varepsilon^{\rm in}_i \varphi^{\rm in}_i (\vec{r}), \,
  {H}_D(t_{\rm out}) \, \varphi^{\rm out}_i (\vec{r})=\varepsilon^{\rm out}_i \varphi^{\rm out}_i (\vec{r}).
\end{equation}
In the final expressions, we will assume that 
$t_{\rm in} \rightarrow -\infty$ and $t_{\rm out} \rightarrow \infty$.
The expected number of electrons $n_k$ in the state~$k$ and the number 
of positrons $\overline n_p$ in the state $p$ are given by~\cite{Greiner_85, Fradkin_91}:
\begin{equation}
  n_ k = \sum_{i<F} \arrowvert a_{ki} \arrowvert^2 \, ,
 \label{eq:electrons}
\end{equation}
\begin{equation}
  \overline{n}_p= \sum_{i>F} \arrowvert a_{pi} \arrowvert^2.
 \label{eq:positrons}
\end{equation}
Here $F$ is the Fermi level ($\varepsilon_F = - m_e c^2$) and
\begin{equation}
 a_{ij} = \int d^3 \vec{r} \, \psi^{(-)\dagger}_i(\vec{r}, t) \, \psi^{(+)}_j (\vec{r}, t)
 \label{eq:tr_ampl}
\end{equation}
are the one-electron transition amplitudes which are time-independent due 
to unitarity 
of the time evolution. The total number of created pairs $P_{\rm t}$
and the number of bound-free pairs $P_{\rm b}$ can be found as
\begin{equation}
  P_{\rm t}=\sum_{k>F} \, n_k = \sum_{p<F} \, \overline{n}_p
  \label{eq:p_t}
\end{equation}
and
\begin{equation}
 P_{\rm b} = \sum\limits_{|\varepsilon_k| < m_e c^2} n_k. 
 \label{eq:p_b}
\end{equation}
Since for the considered processes $P_{\rm t}$ and $P_{\rm b}$
are much smaller than unity, in what follows we will
refer to them as ``probabilities''.
In order to obtain the amplitudes $a_{ij}$,
the right-hand side of Eq.~\eqref{eq:tr_ampl}
can be evaluated at any time moment~$t$.
For $t = t_{\rm out}$ one needs to propagate 
the final eigenstates $ \varphi^{\rm out}_i (\vec{r})$
backward in time from $t_{\rm out}$ to $t_{\rm in}$ 
and then project them on the initial
eigenstates $ \varphi^{\rm in}_i (\vec{r})$:
\begin{equation}
 a_{ij}=\int d^3 \vec{r} \,
   \psi^{(-) \dagger}_i(\vec{r}, t_{\rm in}) \,
   \varphi^{(\rm in)}_j(\vec{r}).
\label{eq:tout_tr_ampl} 
\end{equation}
The advantage of the backward time evolution is 
that the calculation of the bound-free probability $P_{\rm b}$ requires 
propagation of the bound states only.
According to Eq.~\eqref{eq:p_t}, to obtain the total
pair-production probability~$P_{\rm t}$ one has to propagate all
the positive-energy or all the negative-energy states. 
In general,  in order to find all the $n_k$ and $\overline n_p$ 
values (see Eqs.~\eqref{eq:electrons} and~\eqref{eq:positrons}), 
one needs to evolve the full set of the in- or out-eigenstates.
However, in our calculations the Hamiltonian 
has the time-reversal symmetry~$\left( H_{\rm D} (t) = H_{\rm D} (-t) \right)$
and the in- and out-eigenfunctions are identical 
(since $t_{\rm in} = -t_{\rm out}$). It allows us to obtain 
the all positron-creation probabilities~$\overline n_p$ as well 
as the electron ones~$n_k$
via propagation of the positive-energy (or negative-energy)
eigenstates only.
\subsection{Two-center Dirac equation in rotating frame}
We consider a slow symmetric collision of two nuclei. 
It is assumed that the nuclei move along the classical Rutherford 
trajectories and they are treated as sources of an external field. 
The electron dynamics is determined by the time-dependent Dirac 
equation~\eqref{eq:time_dirac}. Let us consider this equation in the reference 
frame rotating with the internuclear axis. In this reference frame,
the Dirac Hamiltonian has the following form:
\begin{equation}
   H_{\rm D} (t) = H_0 (t) - \vec{J} \cdot \vec\omega (t),
   \label{eq:rot_ham}
\end{equation}
where $\vec J$ is the operator of electron total angular momentum,
$\vec \omega (t)$ is the angular velocity of the internuclear axis, and
\begin{equation}
 H_0 (t) = c (\vec\alpha \cdot \vec{p}) + V_{\rm TC} (\vec{r}, t) + \beta m_e c^2.
\end{equation}
Here $V_{\rm TC} (\vec{r}, t)$ is the two-center potential of the nuclei:
\begin{equation}
V_{\rm TC}(\vec{r}, t)=V^A_{\rm nucl} \left(\vec{r}-\vec{R}_A (t) \right)
 +V^B_{\rm nucl} \left( \vec{r}-\vec{R}_B (t) \right) \, ,
\end{equation}
the vectors $\vec{R}_A (t)$ and $\vec{R}_B (t)$ denote the nuclear positions. 
In the present work, we use the uniformly charged sphere model for the nuclear 
charge distribution~$\rho_{\rm nucl}$ and the nuclear potential is given by
\begin{equation}
 V_{\rm nucl} (\vec{r})= \frac{e}{4 \pi}\int d^3 \vec{r}'
  \frac{\rho_{\rm nucl} \left( \vec{r}' \right)}{\lvert \vec{r} - \vec{r}' \rvert}.
\end{equation}
Let us introduce the spherical coordinate system~$\left(r, \theta, \varphi \right)$ 
with the origin at the center-of-mass of the nuclei and the internuclear axis as $z$-axis.
Since the potential $V_{\rm TC}$ does not depend on the azimuthal angle~$\varphi$,
the operator $H_0$ is axially symmetric. But the rotational term $\vec J \cdot \vec \omega $
violates this symmetry. However, in head-on collisions, $\omega \equiv 0$ and $H_{\rm D} (t) = H_0 (t)$.
Moreover, even for collisions with the nonzero impact parameter one can assume that the influence
of the rotational term is not significant and approximate $H_{\rm D} (t)$ by $H_0 (t)$. 
The advantage of this approximation is that there is no coupling between the 
one-electron states $\psi_m$ of different $z$-projection $m$ of the 
total angular momentum. The wave function~$\psi_m$ can be represented
as
\begin{equation}
  \psi_m(r, \theta, \varphi,t)=
 \frac{1}{r}
 \begin{pmatrix}
    G_1(r, \theta, t) \exp [i(m-\frac{1}{2})\varphi] \\
    G_2 (r, \theta, t)\exp [i(m+\frac{1}{2})\varphi] \\
   i F_1 (r, \theta, t)\exp[i(m-\frac{1}{2})\varphi]\\
   i F_2 (r, \theta, t)\exp[i(m+\frac{1}{2})\varphi]
  \end{pmatrix}.
  \label{eq:m_function}
\end{equation}
After substitution of Eq.~\eqref{eq:m_function} in the Dirac equation~\eqref{eq:time_dirac}
using the approximation $H_{\rm D}(t) \approx H_0 (t)$, one can obtain
\begin{equation}
  i \frac{\partial }{\partial t} \Phi(r, \theta, t) = H_m (t) \, \Phi(r, \theta, t)
  \label{eq:2d_dirac}
\end{equation}
for the function 
\begin{equation}
   \Phi(r, \theta, t) = 
   \begin{pmatrix}
    G_1(r, \theta, t)\\
    G_2(r, \theta, t)\\
    F_1(r, \theta, t)\\
    F_2(r, \theta,t)
 \end{pmatrix}.
\end{equation}
Here the operator~$H_m(t)$ is given by
\begin{equation}
   H_{m} (t) =
    \begin{pmatrix} \displaystyle
      m_e c^2 + V_{\rm TC}(t) \quad & c \, D_m \\
      - c \, D_m & -m_e c^2 + V_{\rm TC}(t)
    \end{pmatrix} ,
\end{equation}
where
\begin{equation}
\begin{split}
 D_m = \left( \sigma_z \cos \theta + \sigma_x \sin \theta \right)
 \left( \frac{\partial}{\partial r} - \frac{1}{r} \right)\\
 + \frac{1}{r} \left(\sigma_x \cos \theta - \sigma_z \sin \theta \right) 
 \frac{\partial}{\partial \theta} \\
 + \frac{1}{r \sin\theta} \left( i m \sigma_y + \frac{1}{2} \sigma_x \right)
\end{split}
\end{equation}
and $\sigma_x$, $\sigma_y$, $\sigma_z$ are the Pauli matrices.
In the case of axially symmetric Hamiltonian, one can propagate
the one-electron eigenstates via solving Eq.~\eqref{eq:2d_dirac}
for each $m$ independently.
\subsection{Basis set}
\label{sub:basis}
In order to solve Eq.~\eqref{eq:2d_dirac}, we expand the 
wave function in a finite basis set:
\begin{equation}
 \Phi(r, \theta, t) = \sum\limits_{n = 1}^N C_n (t) W_n (r, \theta),
 \label{eq:fin_exp}
\end{equation}
where $C_n$ are the expansion coefficients and 
the set of $N$ basis functions~$W_n$ is generated using the dual-kinetic-balance (DKB)
technique for axially symmetric systems proposed in Ref.~\cite{Rozenbaum_14}:
\begin{equation}
  W_n(r, \theta) = \Lambda B_i(r) \tilde B_j(\theta) e_u, 
  \quad i = 1,...,N_r,
  \quad j = 1,...,N_\theta,
  \quad u = 1,...,4.
  \label{eq:bfun}
\end{equation}
Here
\begin{equation}
  \Lambda = 
  \begin{pmatrix}
   1 & - \frac{1}{2 m_e c}D_{m}\\
   - \frac{1}{2 m_e c}D_{m} & 1
  \end{pmatrix},
\end{equation}
$ \left\{ B_i(r) \right\}$ and $\{ \tilde B_j (\theta) \}$
are two sets of $N_r$ and $N_\theta$ 
linear-independent one-component functions, correspondingly;
$e_u$ are the unity bispinors; the single index $n \equiv n (i,j, u)$ is 
composed from the indices $i$, $j$, $u$; and $N = 4 N_r N_\theta$.
In our calculation method, 
for $B_i (r)$ and $\tilde B_j (\theta)$ we choose the B-splines 
defined in a spherical box of a finite radius~$L$ 
{with the boundary conditions at $r = L$
set to be zero.} 
The advantage of such choice is that 
the overlap and Hamiltonian matrices are 
sparse. It is due to the fact that only few neighbor spline
overlap. 
It allows us to significantly facilitate the 
numerical calculations.

Substituting the expansion~\eqref{eq:fin_exp} into 
Eq.~\eqref{eq:2d_dirac}, we get
\begin{equation}
 i \sum\limits_{k = 1}^N S_{jk} \frac{d C_k (t)}{dt} = 
 \sum\limits_{k = 1}^N H_{jk} (t)  C_k(t), 
 \label{eq:fin_dirac}
\end{equation}
where $S_{jk}$ and $H_{jk}$ are elements of the overlap and Hamiltonian
matrices, correspondingly:
\begin{equation}
 S_{jk} = \int\limits_0^\pi d\theta \sin\theta
 \int\limits_0^\infty dr \, W_j(r, \theta) W_k (r, \theta)
\end{equation}
and
\begin{equation}
  H_{jk} (t) = \int\limits_0^\pi d\theta \sin\theta
 \int\limits_0^\infty dr \, W_j(r, \theta) H_m (t) W_k (r, \theta).
\end{equation}
Here the integration is performed numerically over the overlap area
of the basis functions. The system~\eqref{eq:fin_dirac} is solved using
the Crank-Nicolson scheme~\cite{Crank_47}:
\begin{equation}
 \begin{split}
 \sum\limits_{ k = 1}^N 
 \left [ S_{jk} \, + \frac{i \Delta t}{2} \,\,  H_{jk}(t+\Delta t/2) \, \right ] \, 
 C_k(t+\Delta t) \,=  \\
 \sum\limits_{ k = 1}^N
 \left[ S_{jk} - \frac{i \Delta t}{2}\, H_{jk}(t+\Delta t/2) \right ] \, C_k(t),
 \label{eq:lin_system}
 \end{split}
\end{equation}
where $\Delta t$ is a sufficiently short time step. This system of linear
equations is solved for each propagated state at each time step employing
the iterative BiCGS (BiConjugate Gradient Squared)
algorithm~\cite{Joly_93} with the preconditioner 
based on an incomplete LU factorization~\cite{Li_11}.

The eigenstates of the instantaneous Hamiltonian 
$H_m (t_{\rm in}) = H_m (t_{\rm out})$ are 
found as the solutions of the generalized eigenvalue problem:
\begin{equation}
 \sum\limits_{k = 1}^N H_{jk} C_k=\sum\limits_{k = 1}^N\varepsilon S_{jk} C_k \, .
 \label{eq:gen_eigen}
\end{equation}
The usage of the DKB technique prevents the appearance of 
the spurious states in the spectrum of Eq.~\eqref{eq:gen_eigen}.
The solutions represent the bound states and the 
both continua. The obtained eigenvectors are propagated in time
according to Eq.~\eqref{eq:lin_system}.
\subsection{Spectrum calculation}
\label{sub:spec}
{
Using a finite basis set, one can calculate the probabilities of 
positron production $\overline n_p$ according to Eq.~\eqref{eq:positrons}.
In order to obtain the energy-differential spectrum~$dP/d\varepsilon$ from the 
discrete set of $\overline n_p$, in Refs.~\cite{Maltsev_15, Bondarev_15},
the Stieltjes method was used}:
\begin{equation}
 \frac{dP}{d\varepsilon} 
 \left(\frac{\varepsilon_p+ \varepsilon_{p+1}}{2} \right)=
 \frac{1}{2} 
 \frac{\overline{n}_{p+1}+\overline{n}_{p}}{\varepsilon_{p+1}-\varepsilon_p},
 \label{eq:stiel}
\end{equation}
{
where $\varepsilon_p$ are the eigenvalues of the Hamiltonian 
matrix~(see Eq.~\eqref{eq:gen_eigen}). 
These calculations were performed in the 
monopole approximation. However, in the two-center case, the resulting 
Hamiltonian matrix exhibits a very nonuniform spectrum with groups of quasidegenerate
eigenvalues. Therefore, some neighboring eigenvalues,  
$\varepsilon_{p+1}$ and $\varepsilon_p$, are very close to each other and 
the corresponding denominator in Eq.~\eqref{eq:stiel} is small enough to
cause the nonphysical resonances in the calculated spectrum. 
It makes impossible 
to use the Stieltjes method. Therefore, 
in the present work, we modify this procedure in the following way}:
\begin{equation}
  \frac{dP}{d\varepsilon} 
 \left(\frac{\varepsilon_p+ \varepsilon_{p+N_S-1}}{2} \right)=
 \frac{1}{\varepsilon_{p+N_S-1}-\varepsilon_p}
 \left(\frac{\overline{n}_{p + N_S-1}+\overline{n}_{p}}{2} + 
 \sum\limits_{i = 1}^{N_S - 2} \overline n_{i + p} \right).
 \label{eq:gen_stiel}
\end{equation}
{
Here $N_S$ is the number of the eigenvalues in the averaging range. 
If $N_S = 2$ then Eq.~\eqref{eq:gen_stiel} is reduced to the simple 
Stieltjes method~\eqref{eq:stiel}. Since, the averaging is performed
over a larger number of points the spurious resonances are smoothed. 
One should choose $N_S$ as small 
as possible in order to prevent oversmoothing of the resulting spectrum.
However, the results obtained according to Eq.~\eqref{eq:gen_stiel} still have
artificial oscillations for any value of $N_S$. 
In order to remove these oscillations, we use the Fourier filtering technique and cut off 
the highest harmonics}:
\begin{equation}
   F_k = \sum\limits_{p  = 0}^{n - 1} J_p \exp \left( -2 \pi i p k / n\right),
  \label{eq:ft}
\end{equation}
\begin{equation}
  J^{\rm cut}_p = 
  \sum\limits_{k  = 0}^{n_{\rm cut} - 1} F_k \exp \left( 2 \pi i p k / n\right).
  \label{eq:inv_ft}
\end{equation}
{
Here} 
\begin{equation}
  J_p = \frac{dP}{d \varepsilon}
  \left(\frac{\varepsilon_p+ \varepsilon_{p+N_S-1}}{2} \right)
\end{equation}
{
are the $n = N/2 - N_S + 1$ ($N$ is the size of the basis set) initial values of the energy-differential spectrum calculated
according to Eq.~\eqref{eq:gen_stiel}, 
$J^{\rm cut}_p$ are the filtered values.
The expression~\eqref{eq:ft}
defines the discrete Fourier transformation, Eq.~\eqref{eq:inv_ft} 
defines the inverse transformation, but summation runs only over 
the $n_{\rm cut} < n$ terms and, therefore, the 
highest $n - n_{\rm cut}$ harmonics are cut from the resulting spectrum.
}

\section{RESULTS}
\label{sec:results}
In this section, we present our results for pair-production probabilities
calculated beyond the monopole approximation. The calculations were performed
for collisions of two bare uranium nuclei moving along
the classical Rutherford trajectories at energy $E = 740$~MeV 
which is near the Coulomb barrier. 
A part of the trajectory with equal
initial and final internuclear distances
$\left( R(t_{\rm in}) = R(t_{\rm out}) = 2 R_0 \right)$ 
was considered. The present results were obtained with 
$R_0 = 250$~fm. 

{The rotation of the internuclear axis was not 
taken into account, i.e., 
the rotational term in Eq.~\eqref{eq:rot_ham} was neglected.}

The basis set was constructed according to Eq.~\eqref{eq:bfun} from
the B-splines of the fourth order in a spherical 
box of size~$L = 10^5$~fm. 
The $\theta$-splines were uniformly distributed
in the range~$[0, \pi]$. The number of $r$-splines $N_r$ was
divided into two parts, $N_r^{(1)}$ and $N_r^{(2)}$. The first part 
was uniformly distributed in the range~$[0, R_0]$.
The last~$N_r^{(2)}$ $r$-splines were placed with 
exponentially increasing step from $r = R_0$ to the border of the box.
It was found that this distribution provides better convergence than
the pure exponential grid. We used the basis set with the following parameters:
$N_\theta = 15$, $N_r = N_r^{(1)} + N_r^{(2)} = 200$,
$N_r^{(1)} = 125$, $N_r^{(2)} = 75$. The generated positive-energy
eigenstates with the energy up to 80~$m_e c^2$, which for this basis
set include 250 bound and 2158 continuum ones, were propagated in order
to obtain the one-electron transition amplitudes.

In Table~\ref{tab:pair_numbers}, we present the obtained results for
probabilities of pair production
for the different values of the impact 
parameter~$b$. The results for the total~$P_{\rm t}$
and bound-free $P_{\rm b}$ pair-production probabilities are
compared with the corresponding values from Ref.~\cite{Maltsev_15}
calculated in the monopole approximation. In the two-center case, 
in contrast to the monopole one,
it is also possible to separate the contribution of 
the quasimolecular ground state, $P_{\rm g}$. 
As one can see from the table, the difference
between the results for $P_{\rm t}$ is about 7\% for $b = 0$
and steadily increases with increasing the value of the impact parameter
reaching 70\% for $b = 40$~fm. This can be explained by the fact that the 
monopole potential better approximates the two-center one
at short internuclear distances. The difference between 
$P_{\rm b}$ values is less than between $P_{\rm t}$ ones and grows slower 
with increasing $b$. It means that, in the two-center case, 
the relative contribution of the
free-free pairs $(P_{\rm f} = P_{\rm t} - P_{\rm b})$ is larger than the corresponding
monopole-approximation contribution, and 
it increases with increasing the impact parameter.
For $b = 40$~fm the two-center free-free probability 
is of the same order of magnitude as the bound-free one, in contrast
with the monopole approximation. It leads to the conclusion 
that effects beyond the monopole approximation have significant influence on the 
free-free pair production, especially for the larger 
values of the impact parameter.
However, the bound states are still the dominant
channel. Moreover, as it follows from Table~\ref{tab:pair_numbers},
the major contribution comes from the pairs with an electron in 
the ground state. 

\begin{table}[h]
\caption{Pair-production probability in 
the U$-$U collision at energy $E$ = 740~MeV 
as a function of the impact parameter $b$. 
$P_{\rm t}$ is the total probability,
$P_{\rm b}$ is the probability of bound-free pair production,  
and $P_{\rm g}$ is the probability of pair production with an electron
captured
into the ground state of the quasimolecule.}
\label{tab:pair_numbers}
\begin{center}
  \begin{tabular}{|c|c|c|c|c|c|}
  \hline
  & \multicolumn{3}{|c|}{Two-center potential} &
  \multicolumn{2}{|c|}{Monopole approximation}\\[0mm]
  \hline
  $b$~(fm)~&$P_{\tiny \rm g}$& \hspace{7mm}  $P_{\tiny \rm b}$  \hspace{7mm} &
  \hspace{7mm} $P_{\rm t}$ \hspace{7mm} & \hspace{7mm}  $P_{\rm b}$  \hspace{7mm} &
  \hspace{7mm} $P_{\rm t}$  \hspace{7mm} \\[1mm]
  \hline
  0   & 1.09 $\times$ $10^{-2}$ &  1.32 $\times$ $10^{-2}$  &  ~~1.38 $\times$ $10^{-2}$~~ & 1.25 $\times$ $10^{-2}$ &  1.29 $\times$ $10^{-2}$~~ \\[0mm]
  5   & 9.3  $\times$ $10^{-2}$ &  1.12 $\times$ $10^{-2}$  &    1.16 $\times$ $10^{-2}$   & 1.05 $\times$ $10^{-2}$ &  1.08 $\times$ $10^{-2}$~~ \\[0mm]
  10  & 6.47 $\times$ $10^{-3}$ &  7.64 $\times$ $10^{-3}$  &    8.01 $\times$ $10^{-3}$   & 7.03 $\times$ $10^{-3}$ &  7.26 $\times$ $10^{-3}$~~ \\[0mm]
  15  & 4.21 $\times$ $10^{-3}$ &  4.87 $\times$ $10^{-3}$  &    5.15 $\times$ $10^{-3}$   & 4.39 $\times$ $10^{-3}$ &  4.51 $\times$ $10^{-3}$~~ \\[0mm]
  20  & 2.73 $\times$ $10^{-3}$ &  3.07 $\times$ $10^{-3}$  &    3.46 $\times$ $10^{-3}$   & 2.70 $\times$ $10^{-3}$ &  2.75 $\times$ $10^{-3}$~~ \\[0mm]
  25  & 1.72 $\times$ $10^{-3}$ &  1.93 $\times$ $10^{-3}$  &    2.14 $\times$ $10^{-3}$   & 1.66 $\times$ $10^{-3}$ &  1.69 $\times$ $10^{-3}$~~ \\[0mm]
  30  & 1.11 $\times$ $10^{-3}$ & 1.23 $\times$ $10^{-3}$  &    1.42 $\times$ $10^{-3}$   & 1.03 $\times$ $10^{-3}$ &  1.04 $\times$ $10^{-3}$~~ \\[0mm]
  40  & 4.72 $\times$ $10^{-4}$ & 5.21 $\times$ $10^{-4}$  &    7.04 $\times$ $10^{-4}$   & 4.09 $\times$ $10^{-4}$ &  4.12 $\times$ $10^{-4}$~~ \\[0mm]
  \hline
  \end{tabular}
\end{center}
\end{table}
It should be noted there exists a little difference 
in $P_{\rm b}$ values with our previous work~\cite{Maltsev_17}. 
This is due to a better accuracy achieved
in the present calculations. 
We also would like to note that the value of $P_{\rm b} = 1.32$ obtained
for $b = 0$ is close to the corresponding one, $P_{\rm b} = 1.29$, from 
Ref.~\cite{RV_Popov_18} calculated using the 
multipole expansion of the two-center potential. 

The results presented in Table~\ref{tab:pair_numbers}
were calculated for the fixed projection~$m = 1/2$ of the 
electron total angular momentum on the $z$ axis 
and then were doubled
to take into account the channel with $m = -1/2$. 
The contribution due to the rotation of the internuclear axis 
was neglected.
In order to investigate the contributions of the 
higher projections~$m$, we calculated the 
probabilities of pair production in the head-on collision 
for $|m| = 3/2$ and 
$|m| = 5/2$. Since there is no rotational coupling in the
head-on collision, the states with different values of $m$
were propagated independently. The results are 
presented in Table~\ref{tab:m}.
As one can see from the table, all the probabilities rapidly
decrease with increasing of $|m|$.
\begin{table}[h]
\caption{Pair-production probability in 
the head-on U$-$U collision at energy $E$ = 740~MeV 
as a function of the absolute value of the angular 
momentum projection~$|m|$. $P_{\rm t}$ is the total probability and
$P_{\rm b}$ is the probability of bound-free pair production.}
\label{tab:m}
\begin{center}
  \begin{tabular}{|c|c|c|}
   \hline
   $|m|$                 &~$P_{\rm b}$~& ~$P_{\rm t}$~\\
   \hline
    ~~$\rfrac{1}{2}$ ~~ & 1.32 $\times$ $10^{-2}$   &   1.38 $\times$ $10^{-2}$ \\
    ~~$\rfrac{3}{2}$ ~~ &  3.50 $\times$ $10^{-7}$   &   3.66 $\times$ $10^{-5}$   \\
    ~~$\rfrac{5}{2}$ ~~ &  6.07 $\times$ $10^{-9}$   &   5.53 $\times$ $10^{-6}$ \\
   \hline
  \end{tabular}
\end{center}
\end{table}

\begin{figure}[h]
\centering 
\includegraphics[trim=0 0 0 0, clip, width = 0.7\textwidth]{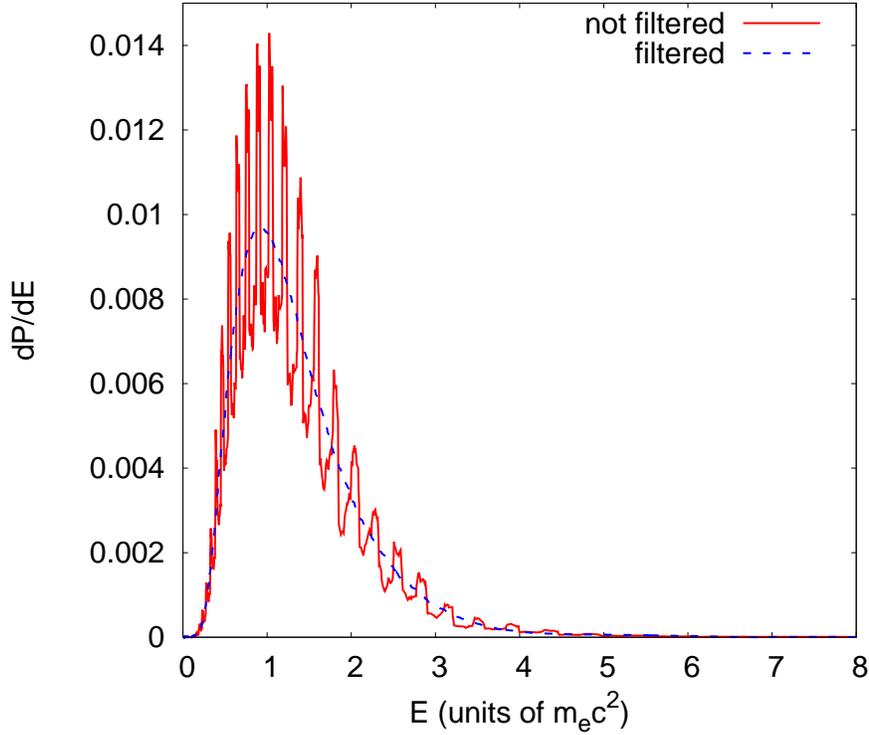}
\caption{Positron energy spectra for the U$-$U head-on collision at energy
$E = 740$~MeV. The dashed (blue) line corresponds to the results obtained with
the filtering procedure and solid (red) one denotes the values obtained without filter.}
\label{fig:filter_work}
\end{figure}

{
The energy spectra of emitted positrons were calculated employing
the method described in  Sec.~\ref{sub:spec}. 
The filtering technique is illustrated in Fig.~\ref{fig:filter_work}
where we present the results obtained using Eq.~\eqref{eq:gen_stiel},
($N_S = 40$) with the Fourier filter and without. The filtering was performed 
according to Eqs.~\eqref{eq:ft} and~\eqref{eq:inv_ft} with $n_{\rm cut} = 70$.
The unfiltered spectrum exhibits many spurious oscillations which occur due 
to the very nonuniform distribution of eigenvalues of the Hamiltonian matrix.
The filtering cuts them off.
}

\begin{figure}[h]
\centering 
\includegraphics[trim=0 0 0 0, clip, width = 0.7\textwidth]{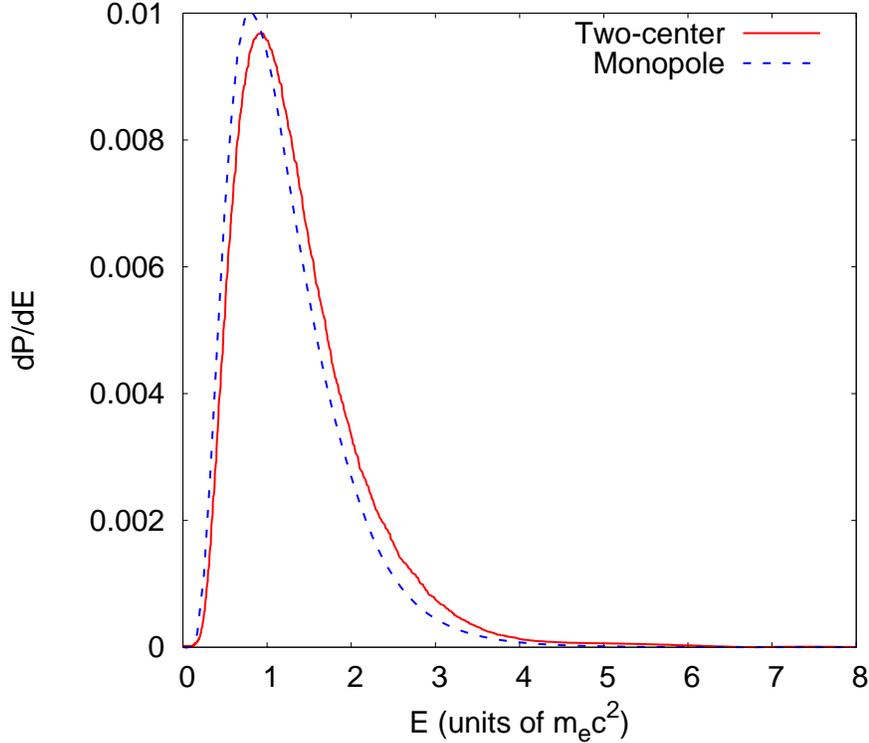}
\caption{Positron energy spectrum for the U$-$U head-on collision at energy
$E = 740$~MeV.
The solid (red) line corresponds to the results obtained with
the full two-center potential and the dotted (blue) one denotes 
the values obtained in Ref.~\cite{Maltsev_15} using 
the monopole approximation.
}
\label{fig:spec_0}
\end{figure}

\begin{figure}[h]
\centering 
\includegraphics[trim=0 0 0 0, clip, width = 0.7\textwidth]{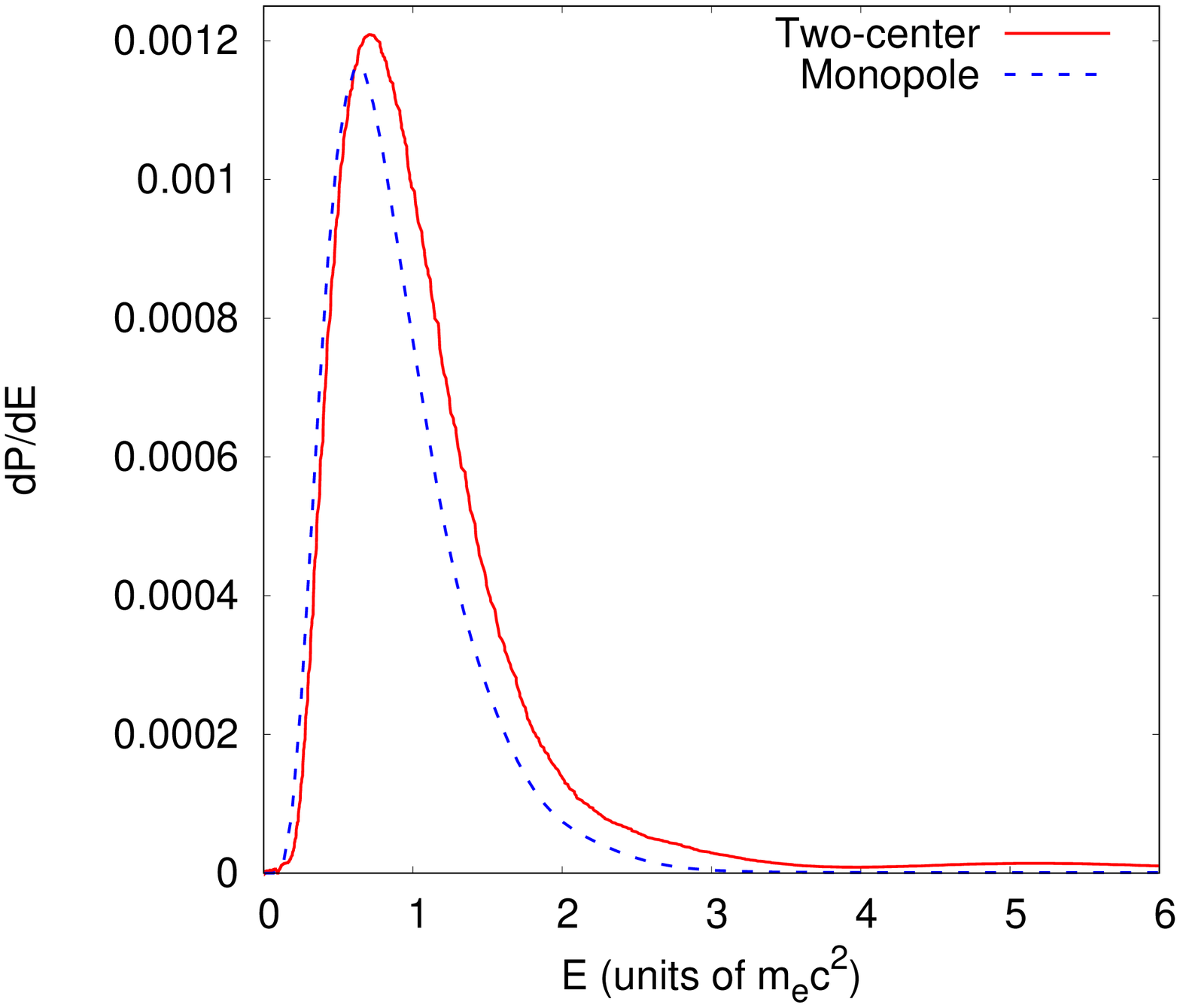}
\caption{Positron energy spectrum for the U$-$U
collision with the impact parameter $b = 30$~fm
at energy $E = 740$~MeV.
The solid (red) line corresponds to the results obtained with
the full two-center potential and the dotted (blue) one denotes 
the values obtained in Ref.~\cite{Maltsev_15} using 
the monopole approximation.
}
\label{fig:spec_30}
\end{figure}

In Figs.~\ref{fig:spec_0} and~\ref{fig:spec_30}, 
we present the obtained positron energy spectra for 
$b = 0$ and $b = 30$~fm. 
The corresponding monopole results from Ref.~\cite{Maltsev_15}
are also shown. It can be seen that the monopole and two-center spectra
are very close to each other. It should be noted that the collision 
with $b = 0$ is supercritical while the collision with $b = 30$~fm is 
subcritical. However, all two-center spectra, as well as the monopole ones,
have the same shape and 
do not exhibit any feature which
can be associated with the spontaneous pair creation or 
the diving phenomenon. 


\section{CONCLUSION}
In the present work, we further evolved our method for calculation of 
pair production in low-energy ion collisions 
beyond the monopole approximation
proposed in Ref.~\cite{Maltsev_17}. Now this technique
allows us to calculate the total pair-production probabilities,
including the free-free ones, and the positron energy spectra 
in low-energy heavy-ion collisions. 

Using the developed method
we calculated pair production in collisions of bare uranium nuclei
at energy near the Coulomb barrier. The obtained results were
compared with the corresponding values from Ref.~\cite{Maltsev_15}
calculated in the monopole approximation.
It was found that the effects beyond the monopole approximation are significant for 
free-free pair production. However, the bound-free pairs dominate 
in the two-center case as well as in the monopole one. 
For small values of the impact parameter the monopole results for
the total pair-production probability
are quite close to the two-center ones, but the difference increases
with increasing the impact parameter.

The positron energy spectra calculated with the full two-center potential 
are very similar to the monopole ones. They do not exhibit any features that 
can be associated to the spontaneous pair production. 
This observation supports the conclusion of 
Refs.~\cite{Reinhardt_81, Muller_88, Maltsev_15} that no direct 
evidence of the diving phenomenon can be found in the positron
energy spectra. However, the methods beyond the monopole approximation make it possible to study the more detailed characteristics of the process under consideration. 
For instance, only the two-center methods allow one to calculate the angular-resolved energy distribution. Therefore these investigations
open new opportunities for searching the scenarios
for indirect detection of the diving phenomenon.

\section*{\large Acknowledgments}
This work was supported by RFBR (Grants No.~16-02-00334 and No.~16-02-00233), 
RFBR-NSFC (Grants No.~17-52-53136 and No.~11611530684), and
by SPbSU-DFG (Grants No. 11.65.41.2017 and No.~STO 346/5-1).
Y.S.K. acknowledges the financial support of FAIR-Russia Research Center and 
CAS~President International Fellowship Initiative~(PIFI) .
The work of V.M.S. was also supported by the CAS President International Fellowship Initiative 
(PIFI) and by SPbSU (COLLAB 2018: No. 28159889). I.A.M. and R.V.P. also
acknowledge support from TU Dresden via the DAAD Programm Ostpartnerschaften.
The research was carried out using computational resources provided by Resource 
Center "Computer Center of SPbSU"

\end{document}